\documentclass{emulateapj}

\usepackage{natbib,graphicx,epsfig}

\newcommand{\psim}{\lower.5ex\hbox{$\; \buildrel \propto \over\sim \;$}}

\shorttitle{Fermi Observations and UHECR 
Acceleration}
\shortauthors{Dermer \& Razzaque}

\received{April 23, 2010}

\begin{document}

\title{Acceleration of Ultra-High Energy Cosmic Rays in the Colliding Shells of Blazars and GRBs: 
Constraints from the Fermi Gamma ray Space Telescope}

\author{Charles D. Dermer$^1$ \& Soebur Razzaque$^{1,2}$ } 

\affil{$^1$Space Science Division,
Code 7653, Naval Research Laboratory,\\ Washington, DC 20375-5352, USA}
\affil{$^2$NRL/NRC Research Associate}

\email{charles.dermer@nrl.navy.mil}


\begin{abstract}
Fermi Gamma ray Space Telescope measurements of spectra,  variability time scale, and maximum photon energy 
give lower limits to the apparent jet powers and, through $\gamma\gamma$ opacity arguments, 
the bulk Lorentz factors of relativistic jets.  The maximum cosmic-ray
particle energy is limited by these two quantities in Fermi acceleration scenarios. 
Recent data are used to
constrain the maximum energies of cosmic-ray protons and Fe nuclei accelerated in colliding shells of 
GRBs and blazars. The Fermi results indicate that Fe rather than protons
are more likely to be accelerated to ultra-high energies in AGNs, whereas
powerful GRBs can accelerate both protons and Fe to $\gtrsim 10^{20}$ eV.
Emissivity of nonthermal radiation from radio galaxies and blazars is estimated from the First Fermi AGN Catalog, 
and shown to favor BL Lac objects and FR1 radio galaxies over flat spectrum radio quasars,  
FR2 radio galaxies, and long-duration GRBs as the sources of UHECRs. 
\end{abstract}

\maketitle
		
\vspace{3mm}

\section{Introduction}

\citet{hil84} pointed out an essential requirement for acceleration of 
ultra-high-energy cosmic rays
(UHECRs), namely that the particle Larmor radius $r_{\rm L}\cong E/QB$
must be smaller than the size scale of the acceleration region. Here 
$E$ is the particle energy, $Q=Ze$ is its charge, and $B$ is the 
magnetic field in the acceleration zone. This limitation
applies to Fermi acceleration scenarios where a particle gains 
energy while diffusing through a magnetized region. Additional limitations 
due, for example, to radiative losses or available time,
 further restrict the maximum energies and therefore the
allowed sites of UHECR acceleration.

Two plausible classes of astrophysical accelerators of 
extragalactic UHECRs are active galactic nuclei (AGN) \citep{mb89,bgg02}
and gamma-ray bursts (GRBs) \citep{wax95,vie95,mu95}
\citep[see also][]{rm98,hh02,dm09},
though other types of sources, including young, highly magnetized neutron stars \citep{ghi08} and 
structure formation shocks \citep{ino08} remain viable. 
The announcement by the \citet{Auger07} of anisotropy in the arrival directions 
of cosmic rays with  energies $E \gtrsim 6\times 10^{19}$ eV, 
even given the reduced correlation in the latest data from the Pierre Auger Observatory
\citep{abr09}, is compatible with the production of UHECRs
in many source classes, including GRBs and blazars. 
Because of the GZK effect involving photohadronic interactions 
of protons or ions  with CMB radiation \citep{gre66,zk66,ste68}, higher-energy cosmic rays with 
$E\gtrsim 10^{20}\,\rm eV$ must be produced by sources located within
distances $d\lesssim 100$ Mpc in order to reach us 
without losing significant energy \citep[e.g.,][]{nw00,hmr06}. 

The most powerful AGNs and long-duration GRBs are found far outside the GZK radius,
at redshifts $z\gtrsim 0.1$. It is therefore of interest to re-examine Fermi
acceleration requirements to determine if there are
AGN and GRB sources with appropriate properties within the GZK radius. Here we make a detailed examination to 
justify a simple derivation of maximum particle energy
relating apparent source power and bulk Lorentz factor $\Gamma$ in the framework of 
Fermi acceleration in colliding shells. (Note that 
these arguments do not apply to non-Fermi type mechanisms, for example, 
 electrodynamic acceleration in pulsar magnetospheres.)
The derived limits are compared
with values implied by Fermi data, yielding constraints on UHECR acceleration in these sources. 
We then use the First Fermi Large Area Telescope (LAT) AGN Catalog (1LAC) \citep{abd10a} to estimate the nonthermal 
emissivity of AGNs. 

We find that the lower luminosity BL Lac objects and FR1 radio 
galaxies are more likely to be the sources of UHECRs than the rare, powerful flat spectrum
radio quasars (FSRQs) and FR2 radio galaxies, and are more likely to accelerate 
Fe than protons to ultra-high energies. GRBs, on the other hand, can accelerate both protons and 
Fe nuclei to ultra-high energies, but are rare within the GZK volume.

\section{Maximum Particle Energy in Colliding Shells}

The total comoving energy 
density $u^\prime$, including rest-mass and magnetic-field energy density, 
of a cold, isotropic relativistic wind with total wind power $L$ and 
outflow Lorentz
factor $\Gamma = 1/\sqrt{1-\beta^2}$ at radius $r$
from the source is
$u^\prime = L/(4\pi r^2 \beta \Gamma^2 c)$. Primes here and 
below refer to quantities measured in the proper (comoving) frame 
of the radiating fluid. If a fraction $\epsilon_B$ of
the total energy density is in the form of 
magnetic-field energy density $u^\prime_{B^\prime} = B^{\prime 2}/8\pi$, where 
$B^\prime$ is the  magnetic field in the fluid frame, then $rB^\prime \Gamma = 
\sqrt{2\epsilon_B L/ \beta c}$, implying maximum particle energies 
$E_{max} \cong \beta \Gamma QB^\prime r^\prime\cong \beta QB^\prime r$ (since the 
comoving size scale $r^\prime \cong r/\Gamma$). Thus
\begin{equation}
E_{max} \cong \left({Ze\over \Gamma}\right) \sqrt{{2\beta\epsilon_B L\over c}}
\cong 2\times 10^{20} Z{ \sqrt{ \epsilon_B \beta L_{46}/\epsilon_e}\over \Gamma}\;{\rm eV}\;,\;
\label{Emax}
\end{equation} 
where the nonthermal $\gamma$-ray luminosity 
$L_\gamma = 10^{46}L_{46}{\rm~erg}{\rm~s}^{-1}$ \citep[e.g.,][]{wax04,fg09}, 
and we write  $L_\gamma = \epsilon_e L$, 
where $\epsilon_e$ is the fraction of jet energy in electrons that
is assumed to dominate the radiative $\gamma$-ray output.  In general, 
$L_\gamma < L$, and $L_\gamma\ll L$ for radiatively inefficient flows.
Besides giving the minimum apparent isotropic jet power, 
$\gamma$-ray observations give minimum values of $\Gamma$ from $\gamma\gamma$ 
opacity arguments, allowing us
to identify whether a given source is a plausible site for UHECR acceleration.

The estimate in eq.\ (\ref{Emax}) does not, however,
explain how a cold magnetohydrodynamic wind
can transform directed kinetic energy to relativistic particles, which
requires consideration of a specific model.
Within the colliding shell framework \citep{rm94,pir99}, 
which is often invoked to explain the formation of spectra in GRBs and blazars,
we can assess the conditions under which eq.\ (\ref{Emax}) is valid.
In this model, a central black hole is assumed to eject
a slower shell $a$ with coasting Lorentz factor $\Gamma_0 = \Gamma_a$ during
 explosion frame times 
$0\leq t_* \leq \Delta t_{*a}$, 
followed by a faster shell $b$ with $\Gamma_0 = \Gamma_b > \Gamma_a$ ejected at 
times $t_{*d} \leq t_* \leq t_* + \Delta t_{*,b}$, where $t_{*d}$ is the stationary-frame
delay time between the start of the ejections of shell a and b.
The shell energies ${\cal E}_{a(b)}$ are related to their luminosities $L_{a(b)}$ through 
${\cal E}_{a(b)} = L_{a(b)} \Delta t_{*a(b)}$.

The shells are assumed to collide after they reach their coasting phase. Neglecting shell spreading (which 
can be included by renormalizing the shell durations), and assuming that the event takes
place sufficiently quickly so that we can approximate the shell density as constant during the duration 
of the collision, then simple kinematics shows that the collision radius $r_{coll}$ and collision time $t_{*,coll}$
are given, in the limit $\Gamma_a\gg 1$, by
\begin{equation}
r_{coll} = ct_{*,coll} \cong {2c\Gamma_a^2(t_{*d}-\Delta t_{*a})\over 1-\rho_\Gamma^2}\;\;,\;\; \rho_\Gamma \equiv {\Gamma_a\over \Gamma_b} < 1\;.
\label{rcoll}
\end{equation}

\begin{figure}[t]
\center
\includegraphics[width= 8.0cm]{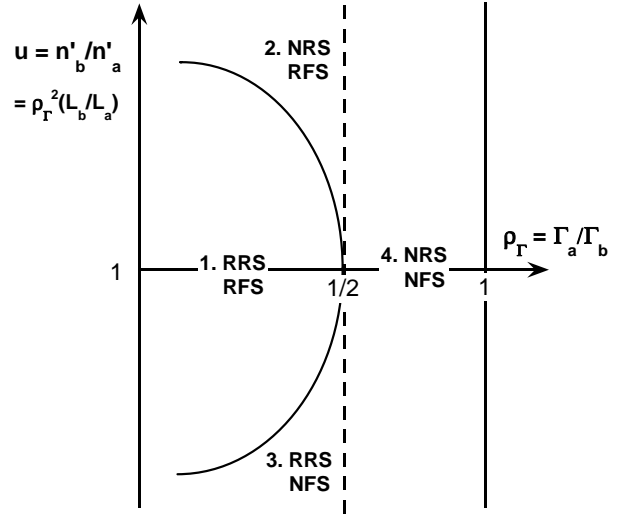}
\caption{
Different regimes in colliding shell interactions.  
}
\label{f1}
\end{figure}

The proper frame particle density in the shells is $n^\prime_{a(b)}= L_{a(b)}/4\pi\Gamma_{a(b)}^2 r^2 m_pc^3$
at radius $r$. Letting $\Gamma(\gg 1)$ denote the Lorentz factor of the shocked fluid, then the Lorentz factors of the 
forward $f$ and reverse $r$ shocks as measured in the shocked fluid frames are $\Gamma_{f(r)} \cong 
\Gamma\Gamma_{a(b)}(1-\beta_{a(b)}\beta) \rightarrow {1\over 2} (\Gamma/\Gamma_{a(b)} + \Gamma_{a(b)}/\Gamma )$
\citep{sp95}. From the equality of kinetic energy densities at the contact discontinuity, we have
\begin{equation}
u\equiv {n^\prime_b\over n^\prime_a} = {\Gamma_f^2 - \Gamma_f\over \Gamma_r^2 - \Gamma_r} = {\Gamma^2_a\over \Gamma_b^2}{L_b\over L_a} = \rho_\Gamma^2 {L_b\over L_a}\;.
\label{u}
\end{equation}

Four asymptotic regimes can be identified, depending on whether the forward shock is relativistic (RFS)
or nonrelativistic (NFS), and the reverse shock is relativistic (RRS) or nonrelativistic (NRS):
\begin{enumerate}
\item RRS ($\Gamma_r \gg 1$) and RFS ($\Gamma_f \gg 1$). This 
holds when $\Gamma_b\gg \Gamma\gg\Gamma_a$, implying 
\begin{equation}
 \Gamma\cong u^{1/4}\sqrt{\Gamma_a \Gamma_b}\;,\; 
\Gamma_f = {u^{1/4}\over 2\sqrt{\rho_\Gamma}}\;,\;{\rm and}\;\Gamma_r = {1\over 2u^{1/4}\sqrt{\rho_\Gamma}}\;
\label{casei}
\end{equation} 
when  $\rho_\Gamma \ll \min(\sqrt{u},1/\sqrt{u})$. 
\item NRS ($\Gamma_r - 1 \approx \beta_r^2/2$) and
RFS, implying 
\begin{equation}
 \Gamma\cong \Gamma_b\;,\; 
\Gamma_f = {1\over 2\rho_\Gamma}\;,\;{\rm and}\;\beta_r = {1\over \sqrt{2u}\rho_\Gamma}\;
\label{caseii}
\end{equation} 
when  $1/\sqrt{2u}\ll \rho_\Gamma \ll 1/2$ or $u\gg 1$. For this case, $L_b \gg L_a$.
\item RRS and NFS ($\Gamma_f - 1\approx \beta_f^2/2)$, implying
\begin{equation}
 \Gamma\cong \Gamma_a\;,\; 
\beta_f = {1\over \rho_\Gamma}\sqrt{u\over 2}\;,\;{\rm and}\;\Gamma_r \cong {1\over 2\rho_\Gamma}\;
\label{caseiii}
\end{equation} 
when  $\sqrt{u/2}\ll \rho_\Gamma \ll 1/2$ or $u \ll 1$. 
\item NRS and NFS,  that is, $\beta_r\ll 1$ and 
$\beta_f \ll 1$, which takes place when $\Gamma_a \cong \Gamma_b \cong \Gamma$. 
 Because $\rho_\Gamma \cong 1$, the duration of the interaction
for this case may be sufficiently long to violate the assumption of constant shell density during the collision, 
though we formally treat it here. 
\end{enumerate}

 Fig.\ 1 illustrates the 
various regimes in terms of the parameters $u$ and $\rho_\Gamma$ 
for which we derive maximum particle energies. 
The magnetic field $B^\prime$ in the shocked fluid is defined in the usual way \citep[e.g.,][]{spn98},
through a magnetic-field parameter $\epsilon_B$, so that 
$B^\prime_{f(r)} = \sqrt{32\pi m_pc^2n^\prime_{a(b)} \epsilon_{Bf(r)}(\Gamma^2_{f(r)}-\Gamma_{f(r)})}$.
The duration of the flare from the forward shock is determined by the time $\Delta t_a^\prime$ required for the 
forward shock to pass through shell $a$. The  
width of shell $a$ in the frame of the shocked fluid is $\Delta_a^\prime/\Gamma_f$ due to length 
contraction, where the proper frame width of shell $a$ is $\Delta_a^\prime = \Gamma \Delta_a(r)$. If $\bar\beta_fc$ 
is the speed of the FS, then $\bar\beta_f \cong 4\beta_f/3$ when $\Gamma_f - 1\ll 1$, and $\bar\beta_f \cong 1$, 
$\bar \Gamma_f \cong \sqrt{2}\Gamma_f$, when $\Gamma_f\gg 1$. The duration of the event is also limited
by the possibility that shell $b$ dissipates its entire energy before the forward shock passes through shell $a$. 
Using similar reasoning for the reverse shock, we obtain the comoving timescales for the interactions at the forward
and reverse shocks:  
$$\Delta t^\prime_{FS(RS)} = \min[ {\Gamma_{a(b)}\Delta_{a(b)}(r)\over \bar\beta_{f(r)} \Gamma_{f(r)} c} ,$$
\begin{equation}
  {{\cal E}_{b(a)}\over 4\pi r^2 n_{a(b)}^\prime \bar \beta_{f(r)}
m_pc^3 (\Gamma_{f(r)}^2-\Gamma_{f(r)})\Gamma}]\;.
\label{Deltatprimea}
\end{equation} 
Following the passage of the reverse and forward shocks through the shells, adiabatic expansion
quickly ends subsequent acceleration and emission \citep{der08}.

The maximum energy of particles accelerated at the forward and reverse shocks is given by 
$E_{max,f(r)} \cong Ze\Gamma B^\prime_{f(r)}c\bar\beta_{f(r)}\Delta t^\prime_{FS(RS)}$.
The general expression can be written as 
$$E_{max,f(r)} \cong {Ze\over \Gamma} \;\sqrt{ {2\epsilon_{Bf(r)}L_{a(b)}\over c} }\big({\Gamma\over \Gamma_a}\big)^2
\sqrt{ {\Gamma_{f(r)}-1\over \Gamma_{f(r)} } }
$$
\begin{equation}
\times { (1-\rho_\Gamma^2 )\min[1, {{\cal E}_{b(a)} \over {\cal E}_{a(b)}}
{\Gamma_{a(b)}\over \Gamma(\Gamma_{f(r)}-1)}]\over (t_{*d} - \Delta t_{*a})/\Delta t_{*a(b)} } \;.
\label{Emaxfr}
\end{equation}
The general expression for the maximum radiative efficiency giving the internal energy 
dissipated in the forward and reverse shocks can be written as
\begin{equation}
{\cal E}_{f(r)} = \min \big[ {\cal E}_{a(b)}{\Gamma (\Gamma_{f(r)}-1)\over \Gamma_{a(b)} }\;,\; {\cal E}_{b(a)}\big]\;.
\label{Efr}
\end{equation}

For the different cases, we obtain the following results:
\subsection{Acceleration at the Forward Shock}
\begin{enumerate}
\item RRS and RFS. 
\begin{eqnarray}
E_{max,f} \cong {Ze\over \Gamma} \sqrt{2\epsilon_{Bf}L_b\over c}\; 
{\min [1, {2{\cal E}_b\over {\cal E}_a}\sqrt{{L_b\over L_a}}]\over (t_{*d} - \Delta t_{*a})/\Delta t_{*a} }\;,\nonumber\\
{\cal E}_f = \min [{\cal E}_b, {{\cal E}_a\over 2}\sqrt{{L_b\over L_a}}]\;.~~~~~~~~~~~~~
\label{Emax1f}
\end{eqnarray}
\item NRS and RFS. 
\begin{eqnarray}
E_{max,f} \cong {Ze\over \Gamma} \sqrt{2\epsilon_{Bf}L_b\over c}\; \sqrt{L_a\over L_b}{{\cal E}_b \over {\cal E}_a} 
{2\min [1, {{\cal E}_a/ 2\rho_\Gamma^2{\cal E}_b}]\over (t_{*d} - \Delta t_{*a})/\Delta t_{*a} }\;,\nonumber\\
{\cal E}_f = \min [{\cal E}_b, {{\cal E}_a/ 2\rho_\Gamma^2}]\;.~~~~~~~~~~~~~
\label{Emax2f}
\end{eqnarray}
\item RRS and NFS. 
\begin{eqnarray}
E_{max,f} \cong {Ze\over \Gamma} \sqrt{2\epsilon_{Bf}L_b\over c}\;  
{\min [1, 4({\cal E}_b/{\cal E}_a)(L_a/L_b]\over (t_{*d} - \Delta t_{*a})/\Delta t_{*a} }\;,\nonumber\\
{\cal E}_f = \min [{\cal E}_a, 4{\cal E}_b(L_a/L_b)]\;,~~~~~~~~~~~~~
\label{Emax3f}
\end{eqnarray}
\item NRS and NFS (also written for acceleration at the reverse shock). 
\begin{eqnarray}
E_{max,f(r)} \cong {Ze\over \Gamma} \sqrt{2\epsilon_{Bf}L_{a(b)}\over c}\times\;~~~~~~~~~~~\nonumber\\
{\beta_{f(r)} (1-\rho_\Gamma^2)\over \sqrt{2}} 
{\min [1, 2{\cal E}_{b(a)}/\beta_{f(r)}^2{\cal E}_{a(b)}]\over (t_{*d} - \Delta t_{*a})/\Delta t_{*a(b)} }\;,\nonumber\\
{\cal E}_{f(r)} = \min [{\beta_{f(r)}^2}{\cal E}_{a(b)}/ 2,{\cal E}_{b(a)}]\;.~~~~~~~~~~~~~
\label{Emax4f}
\end{eqnarray}
\end{enumerate}

\subsection{Acceleration at the Reverse Shock}
\begin{enumerate}
\item RRS and RFS. 
\begin{eqnarray}
E_{max,r} \cong {Ze\over \Gamma} \sqrt{2\epsilon_{Br}L_b\over c}\; 
{2\min [1, {{\cal E}_b/ {2\cal E}_a}]\over (t_{*d} - \Delta t_{*a})/\Delta t_{*a} }\;,\nonumber\\
{\cal E}_r = \min [{\cal E}_a, {{\cal E}_b/ 2}]\;.~~~~~~~~~~~~~
\label{Emax1r}
\end{eqnarray}
\item NRS and RFS. 
\begin{eqnarray}
E_{max,r} \cong {Ze\over \Gamma} \sqrt{2\epsilon_{Br}L_b\over c}\; \sqrt{L_a\over L_b} 
{2\min [1, {{\cal E}_b/{\cal E}_a \over 4u\rho_\Gamma^2}]\over (t_{*d} - \Delta t_{*a})/\Delta t_{*a} }\;,\nonumber\\
{\cal E}_r = \min [{\cal E}_a, {{\cal E}_b/ (2u\rho_\Gamma^2)}]\;,~~~~~~~~~~~~~
\label{Emax2r}
\end{eqnarray}
noting that $u\rho_\Gamma^2 \gg 1$ for this case.
\item RRS and NFS. 
\begin{eqnarray}
E_{max,r} \cong {Ze\over \Gamma} \sqrt{2\epsilon_{Br}L_b\over c}\;  
{\min [1, 2{\cal E}_a/{\cal E}_b]\over (t_{*d} - \Delta t_{*a})/\Delta t_{*a} }\;,\nonumber\\
{\cal E}_r = \min [{\cal E}_a, {\cal E}_b/ 2]\;.~~~~~~~~~~~~~
\label{Emax3r}
\end{eqnarray}
\item NRS and NFS, given by eq.\ (\ref{Emax4f}). 

\end{enumerate}

\section{Limitations on UHECR Acceleration}

The maximum particle energy for the various cases is always proportional to the 
umbrella function, eq.\ (\ref{Emax}), derived from elementary principles, 
but multiplied by a coefficient $\lesssim {\cal O}(1)$. 
The ability of a shell collision to accelerate particles to the highest
energies is conditioned on very specific behaviors of the shells, namely that
the second shell is much faster than the first ($\rho_\Gamma \gg 1$), 
and that the time $t_{*d}$ between shell ejections is a small factor larger 
than the duration $\Delta t_{*a}$ of the event forming shell $a$ (as expressed by the term 
$(t_{*d} - \Delta t_{*a})/\Delta t_{*a}$ in the denominators of eqs.\ (\ref{Emax1f}) -- (\ref{Emax3r})). 
The most favorable regime for particle acceleration to the highest energies
occurs for the case of a RRS and RFS when the energies and 
luminosities of the two shells are about equal. This also gives the 
highest radiative efficiencies. The main requirement is a large 
contrast between the Lorentz factors of the two shells \citep{bel00,kp00}.

The highest radiative efficiency coincides with approximately equal energies
and luminosities for the cases of a RRS and RFS, and a RRS and NFS.  
In the case of a NRS and RFS, where $L_b \gg L_a$ is required for 
validity of this asymptote, a much larger energy in shell $b$
than shell $a$ is required for maximum radiative efficiency at the reverse shock, 
as shown by eq.\ (\ref{Emax2r}). Energy dissipation in this case would, however, more likely be
dominated by the forward shock. Kinematic limitations ensure that the radiative efficiency is  poor for 
dissipation at either the forward or reverse shocks for the case of a NRS and NFS, eq.\ (\ref{Emax4f}), 
depending on the precise energies in each of the shells.

Supposing that the engines of GRBs and blazars or, for that matter, microquasars, eject shells with 
such properties \citep[which is necessary in the case of GRBs to explain their high $\gamma$-ray radiative
efficiency in an internal shock scenario; cf.][]{iok06}, 
then we can construct a diagram illustrating the viability of various sources to
accelerate UHECRs. In Fig. 2 we plot apparent luminosity as a function of Lorentz factor 
for the acceleration of $10^{20}$ eV protons (heavy solid curve) and Fe nuclei (heavy dot-dashed curve),
from eq.\ (\ref{Emax}). Note that we plot this equation to nonrelativistic values of $\Gamma$, 
which is outside the regime where we have considered its validity in a colliding shell scenario.
Sources above these curves can in principle accelerate UHECRs.  
The acceleration rate in Fermi scenarios is governed by the Larmor timescale, so that
the acceleration timescale $t^\prime_{acc} = \phi r^\prime/c = \phi E^\prime/ZeB^\prime c$, and 
$\phi \gtrsim 1$. Incidentally, the requirement that $t^\prime_{acc}$ is smaller than
the available time $t_{ava}^\prime = \Gamma t_v/(1+z)$, where $t_{v}$ is the measured
variability timescale, essentially recovers eq.\ (\ref{Emax}) when $\phi = 1$ and
$r \cong \Gamma^2 c t_v/(1+z)$. This shows that eq.\ (\ref{Emax}) is a restatement of the
Hillas condition by relating $B$ to $L$ and $r_{\rm L}$ to $t_v$.

Maximum particle energy is also limited by the requirement that (i) $t^\prime_{acc}$ 
is shorter than the synchrotron energy-loss timescale $t^\prime_{syn}$ \citep[e.g.,][]{gfr83}. We can 
also consider the (less restrictive) condition (ii) $t^\prime_{syn}> t^\prime_{ava}$, so that particles 
do not significantly cool during the available time. 
Writing the comoving magnetic field $u^\prime_B = \epsilon_B L_\gamma/(\epsilon_e 4\pi r^2 \Gamma^2 c$)
and $E_{20} = E/10^{20}{\rm eV}$,
the former constraint becomes
\begin{equation}
{E_{max,i}\over m_pc^2} \lesssim \Gamma^{5/2} \big({A^2\over Z^{3/2}}\big) 
\big({m_p\over m_e}\big)\sqrt{ {6\pi e c t_v\over \phi \sigma_{\rm T} (1+z)}}\;\big({\beta\epsilon_e c\over 2\epsilon_BL_\gamma}\big)^{1/4}\;,
\label{Emaxi}
\end{equation}
where $Am_p$ is the particle mass, implying
\begin{equation}
L_{\gamma,i}({\rm erg~s^{-1}}) \lesssim {2\times 10^{32}{\Gamma}^{10}\over E_{20}^4}
\;{t^2_v({\rm s})\over \phi^2  (1+z)^2} \big({A^8\over Z^6}\big) \;\big({\beta \epsilon_e \over \epsilon_B}\big)\;,
\label{Lgammai}
\end{equation}
For the second case, 
\begin{equation}
{E_{max,ii}\over m_pc^2}\cong 3\pi\big({A\over Z}\big)^4 \big({m_p\over m_e}\big)^3\;{\beta\epsilon_e\over \epsilon_B} \; {(m_ec^2)\over \sigma_{\rm T} L_\gamma }\;{c^2 \Gamma^6 t_v\over 1+z }\;,
\label{Emaxii}
\end{equation}
implying
\begin{equation}
L_{\gamma,ii}({\rm erg~s^{-1}}) \lesssim  {6\times 10^{38}{\Gamma}^{6}t_v({\rm s})\over   E_{20}(1+z)}\;\big({A\over Z}\big)^4 \;\big({\beta\epsilon_e \over \epsilon_B}\big)\;.
\label{Lgammaii}
\end{equation}

The restrictions implied by eqs.\ (\ref{Lgammai}) and  (\ref{Lgammaii}) are shown by the dashed and dotted lines, 
respectively, in Fig.\ 2, for parameters characteristic of UHECR proton acceleration to 
$10^{20}$ eV in blazars and GRBs. Here $\epsilon_e/\epsilon_B = 1, \phi = 10$, 
and $t_{v} = 10^4$ s and 10 ms, and $\Gamma = 10$ and $10^3$, for blazars and GRBs, respectively.
We also plot data for various sources observed with Fermi and ground-based $\gamma$-ray telescopes. 
In all cases except Centaurus A, $\Gamma_{min}$ is derived from $\gamma\gamma$ opacity arguments, with 
the apparent $\gamma$-ray luminosity giving the minimum source luminosity. The inference of 
$\Gamma_{min}$ from $\gamma\gamma$ opacity arguments is model dependent, with 
the determination of $\Gamma_{min}$ dependent on assumptions about
target-photon anisotropy, relationship between variability time and emission region size scale,
photon escape probability, and the dynamic state of the emitting plasma
\citep[e.g.,][]{ack10,gcc08}. Even so, the strong 
$\Gamma$-dependence of comoving photon energy density $u^\prime_\gamma\propto \Gamma^{-6}$ makes it unlikely that
the actual value of $\Gamma_{min}$ differ by more than a factor of $\approx 2$ 
from the value derived through simple $\gamma\gamma$ arguments.

The long-duration
GRB 080916C \citep{abd09a} and the short-duration GRB 090510A \citep{ack10} have $\Gamma_{min} \approx 10^3$
and $L_\gamma\approx 10^{53}$ erg s$^{-1}$.
GRB 090902B, with $\Gamma_{min} \approx 10^3$ and $L_\gamma \approx 10^{54}$ erg s$^{-1}$ between
6 and 13 s after the trigger time \citep{abd09b}, would 
cluster in the same regime.
For 3C 454.3, $\Gamma_{min} \approx 8$ and $L_\gamma \approx 5\times 10^{48}$ erg s$^{-1}$ \citep{abd09c}.
In the case of NGC 1275, $L_\gamma \approx  10^{42}$ erg s$^{-1}$ and the Doppler factor (and therefore 
$\Gamma$) is $\gtrsim 2$ \citep{abd09d}.
For PKS 2155-304, a BL Lac object, we use the results of \citet{fin08} for the giant 
flares of 2006 July \citep{aha07}, which employs a 
synchrotron self-Compton model with $\gamma\gamma$ absorption and various EBL models to derive $\Gamma\approx 100$. 
Note that the absolute jet powers derived there can be much less than the 
apparent jet power that enters into the acceleration constraint defined by HESS measurements of its
apparent isotropic $\gamma$-ray luminosity $L_\gamma \approx 3\times 10^{46}$ erg s$^{-1}$.

Finally, we consider the case of the FR1 radio galaxy Cen A, the only one of the sources shown in 
Fig.\ 1 that is within the 
GZK radius. It is of special interest, of course, because of the clustering of the arrival directions of several UHECRs
towards Cen A \citep{Auger07,mos09,abr09}, and early speculations that it could be a dominant source of 
UHECRs \citep{pf01}. Because its jet is pointed away from our line
of sight, the jet luminosity and $\Gamma$ factor of Cen A can only be indirectly inferred
\citep{kra02}. One way is to assume that
the energy of the  radio lobes are powered by the jets, and use synchrotron theory and lobe dynamics to infer 
total energy and lifetime. Values between $\approx 10^{42}$ -- $10^{43}$ erg s$^{-1}$ are inferred \citep{har09}, with 
jet beaming and episodes of intense outbursts arguably capable of allowing the jet to reach apparent powers 
sufficient to accelerate UHECR protons \citep{der09}.  

Deceleration of relativistic shells by the surrounding medium
  generates a relativistic external forward shock and a
  relativistic/non-relativistic reverse shock.  The time scale for
  deceleration depends on the apparent isotropic kinetic energy of
  the merged shells $E_{\rm k, iso}$, the bulk Lorentz factor
  $\Gamma_0$ and the density of the surrounding medium $n$, 
  given by $t_{\rm dec} \cong (1+z) (3E_{\rm k, iso}/[32\pi m_pc^5 n
    \Gamma_0^8])^{1/3} \approx 1.9 (1+z) n^{-1/3} E_{55}^{1/3}
  \Gamma_{3}^{-8/3}$~s.  The subsequent evolution of the blast wave is
  described by the \citet{bm76} self-similar solutions.
  Acceleration of cosmic rays to maximum energies in the forward shock
  takes place during the deceleration time, and similar to
  eq. (\ref{Emax}) we can write from $t_{\rm dec} = t_{\rm acc}$
\begin{eqnarray}
E_{\rm max} \cong \frac{Ze}{\phi (1+z)} \frac{\Gamma_0^{1/3}}
{2^{17/12}} (9 \pi E_{\rm k, iso}^2 \epsilon_B^3 n m_pc^2 )^{1/6} 
\nonumber \\
\approx 1.4\times 10^{21}  \frac{Z}{\phi (1+z)} n^{1/6} 
\epsilon_B^{1/2} E_{55}^{1/3} \Gamma_3^{1/3}~ {\rm eV}.
\end{eqnarray}
The corresponding constraint on the apparent isotropic kinetic energy
and bulk Lorentz factor to accelerate particles to $10^{20}$ eV is
\begin{equation}
E_{\rm k, iso} \approx 3.4\times 10^{54} 
\frac{\phi^3 (1+z)^3}{Z^3 \epsilon_B^{3/2} n^{1/2} \Gamma_0}
~{\rm erg}.
\end{equation}
This constraint is satisfied by most long-duration GRBs to accelerate both
protons and Fe nuclei, and by FSRQ blazars (depending in detail on the 
energy of the blazar flare and density of decelerating medium) to accelerate 
Fe nuclei.

\section{Discussion}

Fig.\ 2 shows that the short- and long-duration GRBs for which Fermi observations
give both $L_\gamma$ and $\Gamma_{min}$ easily satisfy the 
luminosity requirements to accelerate UHECR protons or ions. 
After considering the specific parameter values that enter into 
eqs.\ (\ref{Lgammai}) and (\ref{Lgammaii}), 
one finds that the additional constraints imposed by the 
synchrotron cooling rate are not severe either for 
blazars or GRBs. One difficulty for arguing 
that GRBs are the sources of UHECRs is their rarity within the GZK radius. 
Only if the intergalactic magnetic field is sufficiently strong ($\sim$ nG with Mpc scales 
for magnetic-field reversals) to disperse the arrival time of the UHECRs, but not so strong to 
erase their inhomogeneous arrival directions, can long and short GRBs be plausible UHECR candidates \citep{rdf09}.
More complicated magnetic field geometries can also relieve this problem \citep{kw08}.
 A further difficulty accompanying the large $\Gamma$ values and correspondingly dilute
comoving photon energy densities implied by the 
Fermi results on GRBs is that photohadronic processes are suppressed. 
Intermediate neutron production with the escape of ultra-high energy neutrons 
that subsequently decay to form UHECRs was proposed as a principal mechanism \citep{ad03}
to circumvent the problem of the escape of UHECR ions. Such escape is problematic in a bursting 
source because the strong flux of ions will generate a shock that causes the ions
to lose energy as they leave the GRB. \citep{mb10}.  Future Fermi observations will reveal whether 
there is a large population of low $\Gamma$-factor GRBs, or if another class of 
GRBs, such as low-luminosity GRBs \citep{mur06,wan07,lia07}, can make the UHECRs.

From the 1LAC catalog \citep{abd10a}, we can make a diagram, Fig.\ 3, of the volume-averaged 
nonthermal $\gamma$-ray luminosity density (or emissivity). Here we use the  
time-averaged 100 MeV -- 100 GeV luminosity measured over eleven months, 
and divide by the volume $4\pi d^3/3$ associated with the proper distance $d$ of 
the individual sources to make a cumulative emissivity for different classes of 
$\gamma$-ray galaxies. The volume-averaged emissivity, unlike the 
source density, is independent of the beaming factor.
The cumulative emissivities are shown separately 
for BL Lac objects, FSRQs, misaligned AGNs, 
and non-AGN star-forming and starburst galaxies, including 
M82 and NGC 253 \citep{abd10b},
as well as NGC 4945 reported in the 1LAC. NGC 4945 is 
classified here as a starburst, though it also contains a Seyfert nucleus. 
The misaligned AGNs consist of 11 sources, including seven 
FR1 radio galaxies and four FR2 radio sources. The FR1 galaxies are 
Cen A, M87, NGC 1275, NGC 6251, NGC 1218 (3C 78), and PKS 0625-35 \citep{abd10a}.
The FR2 objects consist of two radio galaxies, 3C 111 and PKS 0943-76, and two
steep spectrum radio quasars, 3C 207 and 3C 380.

For comparison with the cumulative emissivity, the fiducial luminosity-density
value of $\approx 10^{44}$ erg Mpc$^{-3}$ yr$^{-1}$ that is needed for
classes of sources to energize UHECRs against GZK losses
 \citep{wb99} is shown. What is obvious from Fig.\ 3 is that
FSRQs do not have sufficient emissivity to power the UHECRs under the 
assumption that the $\gamma$-ray luminosity is a good measure of the 
UHECR power. A much larger energy release in UHECRs than $\gamma$ rays 
is possible, but even so, FSRQs are absent within the GZK radius, and 
 FR2 radio galaxies, which are the putative parent population of
FSRQs  under the unification hypothesis \citep{up95}, are only found 
at distances $\gtrsim 100$ Mpc \citep{mos09}. 
Pictor A is the closest FR2 radio galaxy at $z = 0.035$,
and Cygnus A is at  $z = 0.056$;
neither has yet been reported as Fermi LAT sources.
The redshifts of the  
detected FR2 radio
galaxies 3C 111 and PKS 0943-76 are 0.049, and 0.27, respectively. 

The comparison is more favorable for BL Lac objects which, 
as indicated by 
Fig.\ 3, have the necessary 
nonthermal power to energize UHECRs. The ones detected 
at GeV energies are still outside the GZK radius, though
the famous  TeV (and GeV) blazars 
Mrk 421 and Mrk 501 reside, at $\approx 130$ Mpc, just outside it 
\citep[see also][for a discussion of the space density of 
putative UHECR sources]{ts09}. 
The unification hypothesis would then suggest that many FR1 radio galaxies are
 found at closer distances, including
misaligned galaxies detected at $\gamma$-ray energies. Indeed
FR1 galaxies detected at GeV energies are found within the GZK radius, 
as shown by the misaligned AGNs in Fig.\ 3.
Misalignment means that the $\gamma$-ray luminosity when viewed directly along
the jet is probably much larger than the luminosity measured with Fermi.
NGC 1275 has apparent $\gamma$-ray luminosity of $\approx 10^{44}$ erg
s$^{-1}$ and, at a distance of $\approx 75$ Mpc, falls within the GZK radius \citep{abd09d}. 
It is variable at $\gamma$-ray energies, indicating that much 
of its $\gamma$ ray flux is probably associated with a jet. Moreover, it is a compact 
symmetric object, with transient outbursts with durations of $\approx 10^4$ -- $10^5$ yrs
during which conditions are more favorable for 
UHECR acceleration \citep{ht09}.
The variability is not short enough to give $\Gamma_{min}$ from $\gamma\gamma$ 
arguments, though 
modeling results and observations of apparent superluminal motion suggest 
mildly relativistic Lorentz factors. 
By comparison, Cen A has relatively small apparent $\gamma$-ray lobe and core luminosities,
 each amounting to $\approx 10^{41}$ erg s$^{-1}$, but its emissivity is
large due to its proximity. 

\begin{figure}[t]
\center
\includegraphics[width= 12.0cm]{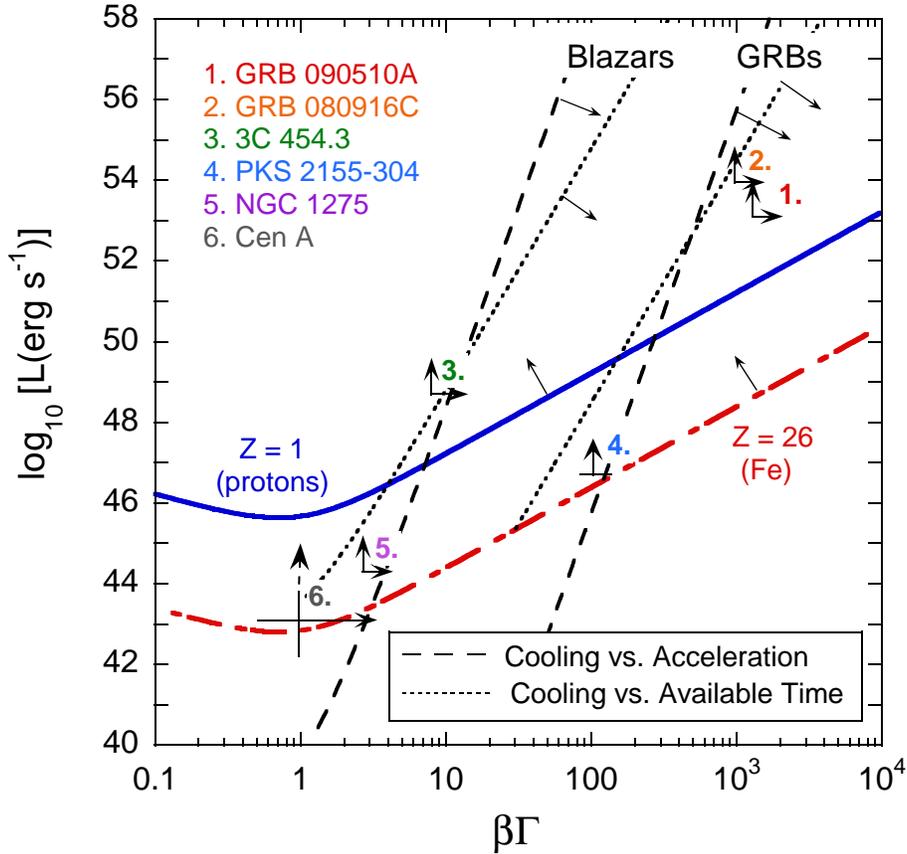}
\caption{
Sources with jet Lorentz factor $\Gamma = \sqrt{\beta^2\Gamma^2+1}$
must have jet power $L$ exceeding the heavy solid and dot-dashed curves to accelerate protons and Fe,
respectively, to $E = 10^{20}$ eV, from eq.\ (\ref{Emax}). Upper limits to $L$ as a function of $\Gamma$ 
for acceleration of UHECR protons to $10^{20}$ eV in blazars and GRBs are given by the dashed lines due to 
competition between synchrotron losses and acceleration, and by the dotted lines when comparing
synchrotron losses and available time. 
Here we use variability time $t_{v} = 10^4$ s  and $z \ll 1 $ for blazars, and 
$t_v = 10$ ms and $z = 1$ for GRBs, as labeled, with $\phi = 10$ in 
both cases. Scalings for different values of $t_{v}$, $\Gamma$,
 $z$, and $\phi$ are given by eqs. (\ref{Lgammai}) and (\ref{Lgammaii}).  
}
\vskip0.1in
\label{f2}
\end{figure}

The star-forming and starburst galaxies are abundant within the GZK radius, 
and have substantial $\gamma$-ray emissivity, which is more than adequate to account
for the power needed to accelerate UHECRs. Where this source class falters, however, is 
in the low individual GeV -- TeV $\gamma$-ray luminosities, 
representing $\approx 3\times 10^{39}$ erg s$^{-1}$ 
for the Milky Way, and $\sim 10^{40}$ erg s$^{-1}$ for  the luminous starbursts \citep{abd10b}. 
Fermi acceleration of UHECRs with such low powers is, as seen from Fig.\ \ref{f2}, not feasible. 
Furthermore, the lack of reported detection of $\gamma$-ray emission from clusters of galaxies
weakens the case for UHECR acceleration in 
structure-formation shocks. 


Long-duration GRBs have apparent $\gamma$-ray luminosities 
far greater than needed to accelerate cosmic-ray protons or ions to $E\gtrsim 10^{20}$ eV,
as shown in Fig.\ 2. Their time-averaged photon 
luminosity density is, however, insufficient to power UHECRs 
within the GZK radius unless the typical baryon loading in GRBs, representing the ratio of energy 
in cosmic rays to that radiated as photons, is $\gg 1$. 
Estimates for the local ($z\ll 1$) luminosity density, based on the 
luminosity function and local event rate density of long-duration GRBs,
range from $\approx 6\times 10^{42}(\Delta t/20{\rm~s})$ erg Mpc$^{-3}$ yr$^{-1}$ 
\citet[][though without using GRB redshift information]{sch01}, 
to $\approx 2\times 10^{44}(\Delta t/10{\rm~s})$ erg Mpc$^{-3}$ yr$^{-1}$
\citep{wp10}, where $\Delta t $ is the mean duration
of long GRBs in the explosion frame. 
A local luminosity density of
($5$ -- $8)\times 10^{42}(\Delta t/10{\rm~s})$
erg Mpc$^{-3}$ yr$^{-1}$ is derived in the treatment of \citet{gpw05}
and the luminosity function 
of \citet{lia07} implies a local luminosity density of
$\approx 2\times 10^{43}(\Delta t/10{\rm~s})$ erg Mpc$^{-3}$ yr$^{-1}$.
Based on a physical model of GRB jets, \citet{ld07} calculate a 
long duration GRB density of ($3$ -- $4)\times 10^{43}$
erg Mpc$^{-3}$ yr$^{-1}$, assuming $\Delta t = 10$ s. 

By comparison with the nonthermal 
luminosity density of long-duration GRBs, that of 
FR1 radio galaxies and BL Lac objects within the GZK radius is at least
1 -- 2 orders of magnitude larger (Fig.\ 3). If the comparison is with
the nonthermal emission emitted in the GeV/LAT range rather than at MeV 
energies, which could be  
thermal or photospheric emission, then the 
required baryon loading must be an order of magnitude larger \citep{eic10}.
The local photon luminosity densities 
of the short hard GRBs \citep[e.g.,][]{gue06} 
or low luminosity GRBs \citep{wan07,lia07,mur08} can also be comparable to the 
emissivity from long-duration GRBs, though with a larger local space density and 
smaller energy release per event.


\section{Conclusions}

Fermi observations shown in Fig.\ 3 indicate that FR1 radio galaxies
and misaligned BL Lac objects located within the GZK radius have 
sufficient emissivity to power the UHECRs. With 
typical Lorentz factors $\approx 2$ -- 10, and apparent jet powers $\approx 10^{44}$ -- $10^{45}$ erg s$^{-1}$
(which could
exceed $10^{46}$ erg s$^{-1}$ and large Lorentz factors during flaring episodes), 
Fig.\ 2 shows that acceleration 
of Fe nuclei in FR1 radio galaxies is possible in colliding shells made in the 
jets of these galaxies. Given the favorable circumstances needed for colliding shells to accelerate
UHECRs, including large Lorentz factor contrast and short times between 
shell ejections, the acceleration of protons is less likely.
The $L-\Gamma$ constraint is unfavorable for UHECR
acceleration at sites with low apparent luminosity, such as starburst
galaxies or the lobes of radio galaxies. 

Long-duration GRBs have sufficient power to accelerate cosmic rays to ultra-high energies,
but their local photon luminosity density in photons, $\sim 10^{43}$ -- $10^{44}$ erg Mpc$^{-3}$ yr$^{-1}$,
implies comparable or large baryon loading
in most models for UHECR acceleration. When compared with the clearly nonthermal 
Fermi LAT flux, the required baryon-loading becomes significant, as shown by \citet{eic10}.
The local nonthermal luminosity density of FR1 radio galaxies and BL Lac objects
by far dominates that of GRBs, especially when compared only with the LAT fluxes from GRBs and blazars.
This circumstance favors UHECR acceleration by 
the supermassive black-hole engines in radio galaxies and blazars, 
provided that UHECRs are predominantly Fe ions.

\acknowledgments

We thank R.\ Blandford, C.\ C.\ Cheung,  D.\ Eichler, and  J.\ Finke for helpful 
discussions and correspondence,
and acknowledge useful comments by the referee.
This work is supported by the Office of Naval Research and NASA 
Fermi Guest Investigator grants NNG 10PK07I and NNG 10PE02I.

\begin{figure}[t]
\center
\vskip0.05in
\includegraphics[width= 12.0cm]{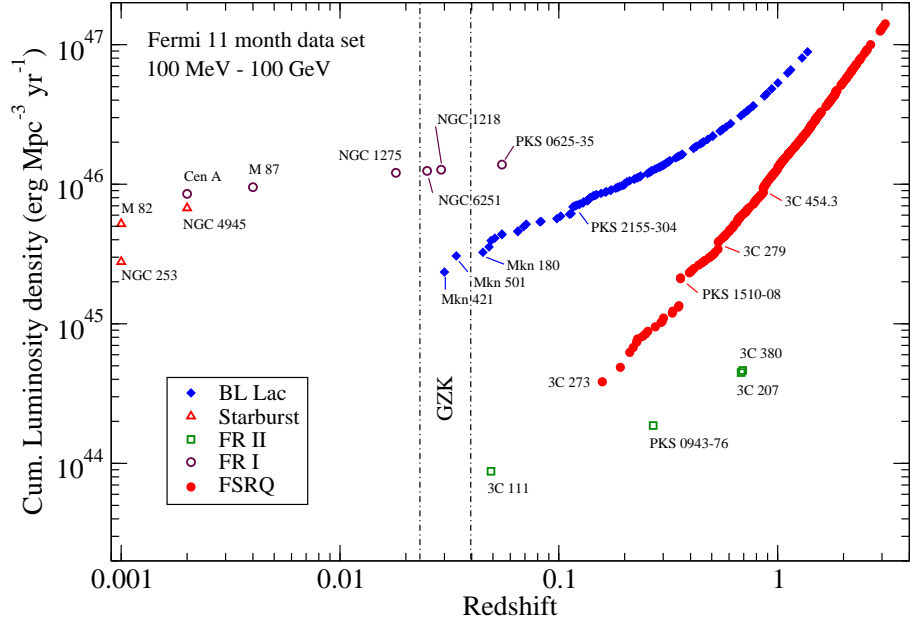}
\caption{
Nonthermal luminosity density of different classes of $\gamma$-ray galaxies detected
with Fermi.  
Here we show cumulative emissivities for FSRQ, BL Lacs, FR1 and FR2 radio galaxies, and star-forming galaxies.
The band between $\approx 100$ and 200 Mpc labeled ``GZK" represents the outer perimeter from which 
sources of UHECRs with energies $\gtrsim 10^{20}$ eV can originate.  
}
\vskip0.2in
\label{f3}
\end{figure}

\end{document}